\documentclass[twocolumn,showpacs,floats,floatfix,superscriptaddress,aps,pra]{revtex4-1}
\usepackage{amsfonts}
\usepackage{amssymb}
\usepackage{amsmath}
\usepackage{calc}
\usepackage{graphicx}
\usepackage{bm}

\usepackage[normalem]{ulem}
\usepackage{amsmath,amssymb}
\usepackage{multirow}
\usepackage{xcolor,soul}
\usepackage{srcltx}
\usepackage{hyperref,graphicx}

\def\be{ \begin{equation}}
\def\ee{ \end{equation}}
\def\bea{ \begin{eqnarray}}
\def\eea{ \end{eqnarray}}
\def\bse{ \begin{subequations}}
\def\ese{ \end{subequations}}
\def\bc{ \begin{center}}
\def\ec{ \end{center}}

\def\H{\mathbf{H}}

\def\I{\mathbf{I}}

\def\T{\mathbf{T}}
\def\D{\mathbf{D}}
\def\M{\mathbf{M}}
\def\V{\mathbf{V}}

\def\N{\mathcal{N}}
\def\Re{\text{Re}\,}

\def\min{\text{min}}
\def\coh{\text{coh}}

\def\ket#1{\vert #1 \rangle}
\def\bra#1{\langle #1 \vert}
\def\braket#1#2{\langle #1 \vert #2 \rangle}

\begin{document}

\author{Boyan T. Torosov}
\altaffiliation{Permanent address: Institute of Solid State Physics, Bulgarian Academy of Sciences, 72 Tsarigradsko chauss\'{e}e, 1784 Sofia, Bulgaria}
\affiliation{Dipartimento di Fisica, Politecnico di Milano and Istituto di Fotonica e Nanotecnologie del Consiglio Nazionale delle Ricerche, Piazza L. da Vinci 32, I-20133 Milano, Italy}
\author{Giuseppe Della Valle}
\affiliation{Dipartimento di Fisica, Politecnico di Milano and Istituto di Fotonica e Nanotecnologie del Consiglio Nazionale delle Ricerche, Piazza L. da Vinci 32, I-20133 Milano, Italy}
\author{Stefano Longhi}
\affiliation{Dipartimento di Fisica, Politecnico di Milano and Istituto di Fotonica e Nanotecnologie del Consiglio Nazionale delle Ricerche, Piazza L. da Vinci 32, I-20133 Milano, Italy}
\title{Quantum simulation of Riemann-Hurwitz $\zeta$ function}
\date{\today}

\begin{abstract}
We propose a simple realization of a quantum simulator of the Riemann-Hurwitz (RH) $\zeta$ function based on a truncation of  its Dirichlet representation. We synthesize a nearest-neighbour-interaction Hamiltonian, satisfying the property that the temporal evolution of the autocorrelation function of an initial bare state of the Hamiltonian reproduces the  RH function along the line $\sigma+i \omega t$ of the complex plane, with $\sigma>1$. The tight-binding Hamiltonian with engineered hopping rates and site energies can be implemented in a variety of physical systems, including trapped ion systems and optical waveguide arrays. The proposed method is scalable, which means that the simulation can be in principle arbitrarily accurate. Practical limitations of the suggested scheme, arising from a finite number of lattice sites $N$ and  from decoherence, are briefly discussed.
\end{abstract}

\maketitle

\section{Introduction}

Riemann $\zeta$ function has attracted the interest of mathematicians and physisists for quite a long time. This function plays a key role in number theory, in particular in the distribution of prime numbers \cite{Euler,Schleichbook}, and it is the basis of one of the most fundamental mathematical conjectures, the Riemann hypothesis. In physics, Riemann's $\zeta$ function is found to be related to a wide variety of different physical areas and phenomena, ranging from classical mechanics to statistical and quantum  physics. For an extensive review of the topic, see \cite{Riemann-RMP}. Several works have highlighted the close connections among the Riemann hypothesis, random matrix theory and  the physics of  classical and quantum chaos (see, for instance, \cite{Berry1,Berry2,PNAS,billard,Bourgade} and references therein). In statistical physics, the Riemann $\zeta$
function can be seen as the partition function of a quantum gas, called the Bose Riemann Gas
with the prime numbers labeling the eigenstates  \cite{Julia,uffa1,uffa2}. In quantum mechanics, several attempts have been 
made to introduce a quantum system whose spectrum is associated with the Riemann $\zeta$ function. In particular, 
following the original Hilbert-P\'{o}lya conjecture great efforts have been devoted to propose quantum Hamiltonians whose bound states coincide with the zeros of the $\zeta$ function \cite{Riemann-RMP,Berry2,kepalle,kepalle1,kepalle2}. Riemann $\zeta$ zeros also enter into phenomena like Bose-Einstein condensation or the cosmic microwave background. For example, a Bose-Einstein condensate could be used, at least in principle, to factorise numbers and to calculate the prime
factors \cite{Holthaus}.\\ In a recent work \cite{Schleich}, Mack and collaborators showed that a generalization of the Riemann function, introduced by Hurwitz \cite{Hurwitz}, can be retrieved from the autocorrelation function of a quantum state  propagating in an anharmonic oscillator potential. The connection to this Riemann-Hurwitz (RH) $\zeta$ function is made through its Dirichlet representation, and requires that the quantum system be initially prepared into a thermal phase state, the so-called Riemann state \cite{Schleich}. Unfortunately, such a thermal phase state is not the mixed thermal state of the anharmonic oscillator, and system preparation might be a nontrivial issue involving a coherent state superposition.\\ 
In this work, we propose a simple method to simulate the RH function by using an $N$ dimensional quantum system, in which system preparation in a thermal phase state is readily provided by excitation of a bare state of the system. Interestingly, our finite-dimensional system is described by a nearest-neighbour-interaction Hamiltonian, which can be implemented in a variety of quantum or classical systems, such as trapped ions or optical waveguide arrays with engineered hopping rates.\\ 
The paper is organized as follows. In Sec.II, we briefly review the connection between the correlation function of a specially-prepared quantum system and the Dirichlet representation of the RH function, introduced in Ref.\cite{Schleich}. In Sec.III we introduce a rather general technique to synthesize a finite-dimensional tridiagonal Hamiltonian satisfying the property that  the temporal evolution of the autocorrelation function of an initial bare state of the system reproduces the RH function along the complex axis $\sigma+i \omega t$, with $\sigma>1$. In Sec.IV we briefly present possible physical implementations of the nearest-neighbor Riemannian Hamiltonian, based on trapped ions or optical waveguide array systems. Finally, in the concluding section V we discuss the main limitations of our quantum simulator scheme and  briefly suggest possible extensions of our results.

\section{Basic idea}
\begin{figure}[tb]
\includegraphics[width=7cm]{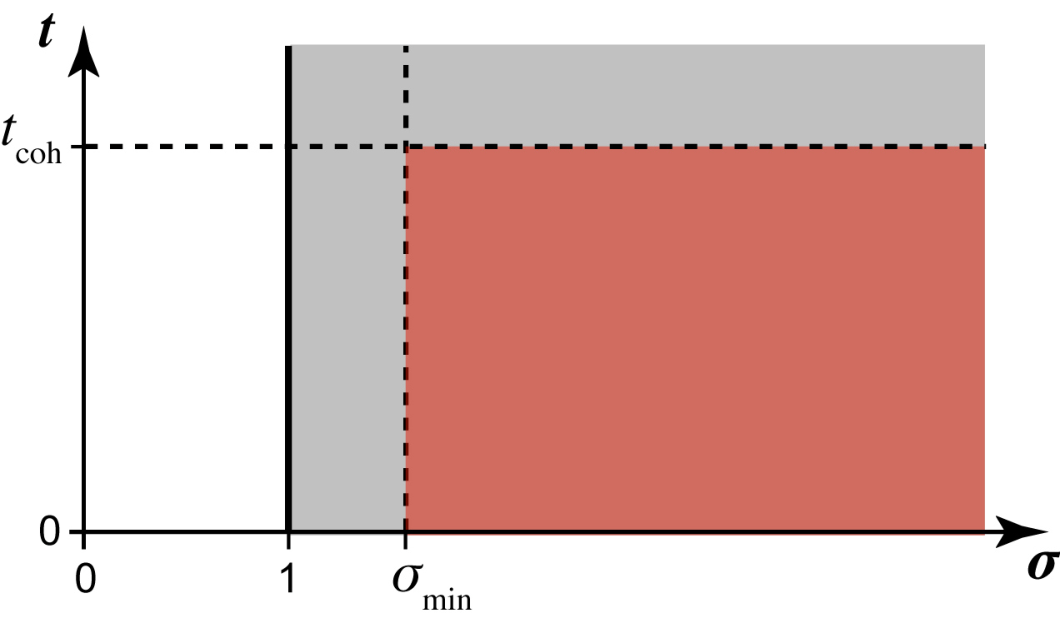}
\caption{Domain of convergence of the series \eqref{Dirichlet} (shadowed light area) in the complex $s=\sigma+i \omega t$ plane. The shaded dark area schematically shows the accessible domain of our quantum simulator of the Riemann $\zeta$ function. The boudaries of this domain arise from the truncation of the Dirichlet series \eqref{Dirichlet} ($ \sigma > \sigma_{\min}$) and from the finite coherence time of the system ($ t<t_{\coh}$). An estimate of $\sigma_{\min}$ is given in Sec.V.}
\label{Region}
\end{figure}

We consider an $N$-dimensional Hilbert space, in which $\ket{0}, \ket{1}, \ldots, \ket{N-1}$ is an orthonormal basis and the Hamiltonian describing our system in such a basis is the $N\times N$ matrix $\H$. Following the idea in \cite{Schleich}, we assume that the Hamiltonian $\H$ has a logarithmic energy spectrum,
\be\label{spectrum}
E_n=\hbar\omega\ln(n+a),
\ee
where $0 < a \leq 1$ and $n=0,1,\ldots, N-1$. Initially, we prepare our system in state $\ket{\psi(0)}=\ket{0}$, which is assumed to be easy to prepare experimentally. Then this state evolves according to
\be
\ket{\psi(t)}=\sum_{n=0}^{N-1} C_n e^{-i E_n t/\hbar} \ket{\alpha_{n}} ,
\ee
where $\ket{\alpha_{n}}$ is the $n$th eigenstate of $\H$ with an eigenvalue $E_n$,
\be
\H \ket{\alpha_{n}} =E_n\ket{\alpha_{n}} ,
\ee
and $C_n= \langle \alpha_{n} | 0 \rangle$. Let us now suppose that the amplitudes $C_n$ can be set to
\be\label{thermal}
C_n = \N (n+a)^{-\sigma /2} ,
\ee
where $\N$ is a normalization factor. If the above assumptions are fulfilled, after some simple algebra, the correlation function, i.e. the probability amplitude of state $\ket{0}$ at time $t$, is given by
\be\label{correlation}
\braket{0}{\psi(t)}=\frac{1}{\N^2}\sum_{n=0}^{N-1}\frac{1}{(n+a)^{\sigma +i\omega t}} .
\ee
At this point, we can compare this expression with the Dirichlet representation of the RH $\zeta$ function, given by
\be\label{Dirichlet}
\zeta(s,a)=\sum_{n=0}^{\infty}\frac{1}{(n+a)^s} 
\ee 
where we have set $s=\sigma+i \omega t$. The Dirichlet series  on the right hand side of Eq.~\eqref{Dirichlet} converges for $\Re s \equiv \sigma >1$, as shown in Fig.~\ref{Region}. An accurate estimate of the RH $\zeta(s,a)$ function can be obtained by truncating the series up to a certain order $N$. This is precisely what our quantum simulator does, see Eq.~\eqref{correlation}. A detailed discussion of the error arising from truncation will be discussed in Sec.V. Note that the ordinary Riemann $\zeta$ function is obtained for $a=1$. This simple connection allows for the simulation of the RH function, as far as the following requirements are met: (i) the Hamiltonian has a logarithmic spectrum \eqref{spectrum}; (ii) the initial amplitudes of the eigenstates of $\H$ obey Eq.~\eqref{thermal}; (iii) there exists a simple experimental setup, providing such Hamiltonian. 
In Fig.~\ref{Region} we show the allowed part of the complex plane, where our approach is applicable. Unfortunately, the most interesting part, $\sigma=1/2$, cannot be accessed. We will return to this point in Sec.V.
In the next section we propose a practical approach, which allows to meet the above requirements for the Hamiltonian.

\section{Construction of the Hamiltonian}
In this section we propose a rather general technique to synthesize a tridiagonal Hamiltonian matrix $\H$ with the required spectrum \eqref{spectrum}, which should also meet the condition \eqref{thermal}. To this aim, we start with the diagonal matrix defined by the logarithmic eigenvalues \eqref{spectrum}, and  we make a sequence of similarity transformations (basis changes), which do not change the eigenvalues but modify the eigenvectors of the Hamiltonian matrix. Let us indicate by $\mathbf{D}$ the  $N$-dimensional diagonal matrix 
\be
\D=\text{diag}\left[ E_0,E_1,\ldots, E_{N-1} \right] ,
\ee
where the energies $E_n$ are assumed to follow the logarithmic dependence \eqref{spectrum}.
Next, let us make a basis transformation, by using an orthogonal matrix $\T$,
\be
\H^{\prime} = \T^{-1} \D \T ,
\ee
where we choose the elements of the first column of $T$ equal to the amplitudes $C_n$ in Eq.~\eqref{thermal}. The other columns are constructed as orthonormal vectors of the orthogonal complement. This task has infinitely many solutions, which, however, are identical regarding our purposes. Finally, we want to transform the matrix $\H^{\prime}$ into a tridiagonal form. This can be achieved, for instance, by using a tridiagonalization procedure, based on Householder reflections \cite{HR}. The described algorithm leads to a tridiagonal Hamiltonian $\H$, which has the desired logarithmic spectrum, and also fulfils condition \eqref{thermal}. For the sake of clarity, we present an example of the described procedure for $N=5$. We choose the parameters in the simulation to be $a=1/2$ and $\sigma = 2$. In this case, the matrix $\D$, according to Eq.~\eqref{spectrum}, is
\be
\D = \hbar\omega\, \text{diag}\left[ \ln(1/2), \ln(3/2), \ln(5/2), \ln(7/2), \ln(9/2) \right] .
\ee
The transformation matrix $\T$ has the approximate numerical value
\be
\T = \left[\begin{array}{ccccc}
0.919 & 0.394 & 0 & 0 & 0 \\
0.306 & -0.714 & 0.629 & 0 & 0  \\
0.184 & -0.429 & -0.576 & 0.671 & 0   \\
0.131 & -0.306 & -0.412 & -0.585 & 0.614 \\
0.102 & -0.238 & -0.320 & -0.455 & -0.789 
\end{array}\right] ,
\ee
where the elements of the first column are given by Eq.~\eqref{thermal}, and the next columns are constructed such that all vector-columns are orthonormal, which leads to an orthogonal matrix, $\T^{-1}=\T^T$. Using this transformation matrix, we obtain for the Hamiltonian in the new basis
\begin{align}\notag
&\H^{\prime}= \T^{-1} \D \T  \\
&=\hbar\omega\left[\begin{array}{ccccc}
-0.479 & -0.499 & -0.136 & -0.053 & -0.020 \\
-0.499 & 0.470 & 0.317 & 0.124 & 0.047 \\
-0.136 & 0.317 & 0.831 & 0.167 & 0.064 \\
-0.053 & 0.124 & 0.167 & 1.153 & 0.090 \\
-0.020 & 0.047 & 0.064 & 0.090 & 1.409
\end{array}\right] .
\end{align}
Finally, we proceed to the tridiagonalization of $\H^{\prime}$. This is achieved by using a sequence of $N-2$ Householder transformations,
\be\label{sequence}
\H=\M(v_3)\M(v_2)\M(v_1)\H^{\prime} \M(v_1)\M(v_2)\M(v_3),
\ee
where
\be\label{QHR}
\M(v)=\I -2\ket{v}\bra{v}
\ee
and
\bse \label{vectors}
\begin{align}
&\ket{v_1}=\left[ 0,0.990,0.132,0.052,0.020 \right]^T, \\
&\ket{v_2}=\left[ 0,0,0.959,0.256,0.119 \right]^T, \\
&\ket{v_3}=\left[ 0,0,0,0.941,0.337 \right]^T
\end{align}
\ese
The explicit numerical value for the Hamiltonian is
\be\label{explicit}
\H=\hbar\omega \left[\begin{array}{ccccc}
-0.479 & 0.520 & 0 & 0 & 0 \\
0.520 & 0.701 & 0.445 & 0 & 0 \\
0 & 0.445 & 0.894 & 0.311 & 0 \\
0 & 0 & 0.311 & 1.062 & 0.198 \\
0 & 0 & 0 & 0.198 & 1.208
\end{array}\right] ,
\ee
where we have neglected the minus signs of the negative off-diagonal elements, since they can be removed by only a simple phase transformation in the amplitudes. Obtained in this way, the Hamiltonian $\H$ has the desired logarithmic spectrum, and if we diagonalize it,
\bse
\begin{align}
&\D=\V^{-1} \H \V, \\
&\V=\left[\begin{array}{ccccc}
0.919 & 0.306 & 0.184 & 0.131 & 0.102 \\
-0.378 & 0.521 & 0.493 & 0.437 & 0.389 \\
0.110 & -0.704 & 0.024 & 0.389 & 0.583 \\
-0.020 & 0.361 & -0.703 & -0.177 & 0.586 \\
0.002 & -0.089 & 0.478 & -0.781 & 0.392
 \end{array}\right],
\end{align}
\ese
we notice that the first row of $\V$ is equal to the first column of $\T$, and is equal to the numbers in Eq.~\eqref{thermal}. The fact that $\T^T$ and $\V$ have the same first row follows from the property that the Householder transformations in Eq.~\eqref{sequence}, with vectors \eqref{vectors}, do not change the first row of the matrix, which diagonalizes the Hamiltonian. Hence the matrices $\T^T$ and $\V$, which diagonalize respectively $\H^\prime$ and $\H$, have the same first row.
The matrix $\T$ is in fact the matrix that connects the two bases $\ket{n}$ and $\ket{\alpha_{n}}$, which leads to the possibility to express state $\ket{0}$ as a coherent superposition of the eigenstates of $\H$, 
\be
\ket{0}=\sum_{n=0}^{N-1} C_n \ket{\alpha_{n}}
\ee
where $C_n$ have the desired values \eqref{thermal}. In such way we have fulfilled all of the necessary restrictions for the Hamiltonian. The described procedure can be applied for an arbitrary dimension $N$, which means that in principle we can simulate RH's $\zeta$ function with an arbitrary accuracy.

\begin{figure}[tb]
\includegraphics[width=\columnwidth]{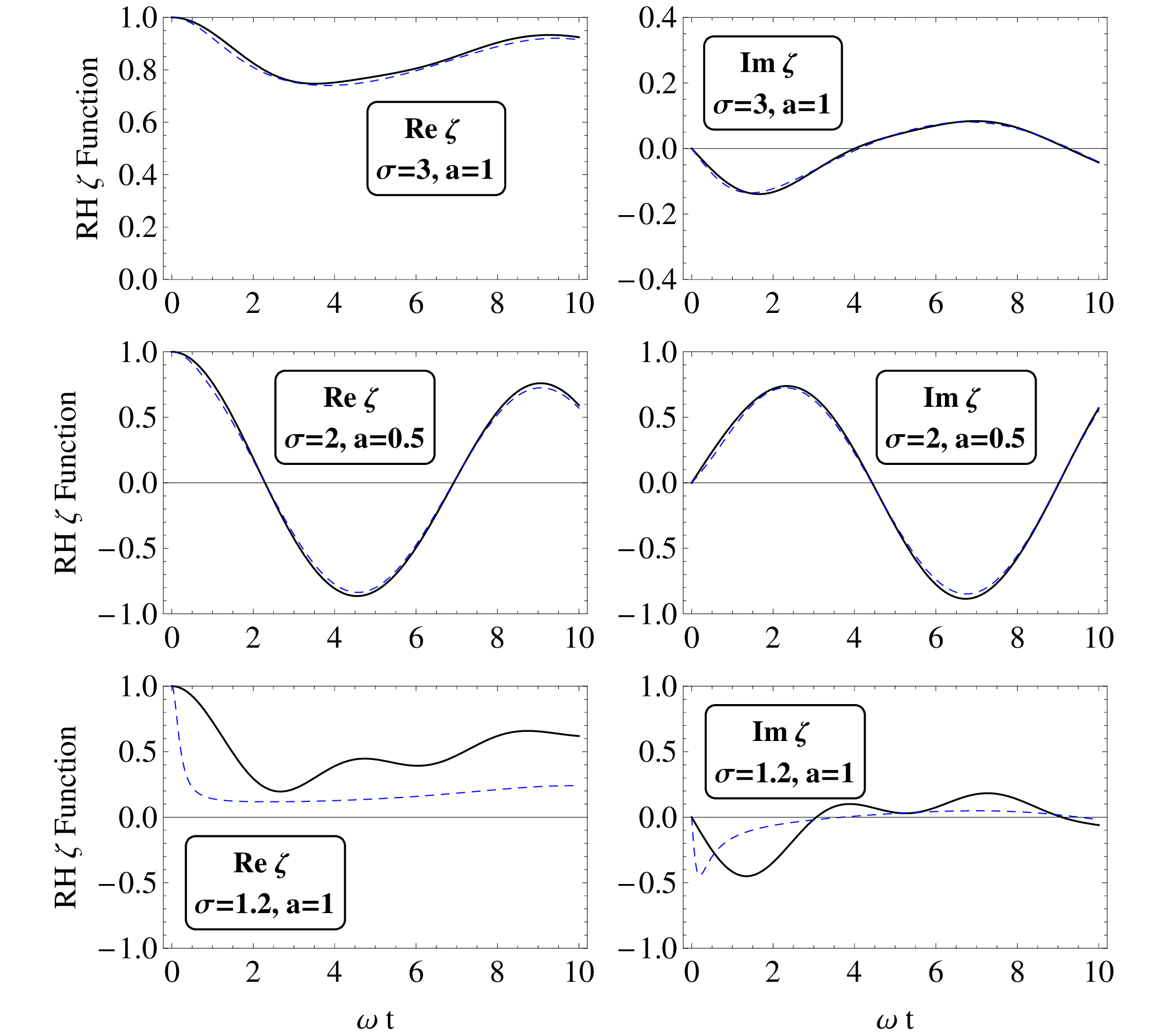}
\caption{Comparison of the normalized RH zeta function $\zeta(\sigma + i\omega t,a)/\zeta(\sigma,a)$ (solid line), with the simulation, obtained by numerical integration of the Schr\"{o}dinger equation, assuming $N=5$ (dashed line). Upper and lower panels: $a=1$ (ordinary Riemann function); middle panels: $a=0.5$.}
\label{comparison}
\end{figure}

As an example, in Fig.~\ref{comparison} we compare the simulation of the RH function, obtained by numerical integration of the Schr\"{o}dinger equation, with the properly normalized RH $\zeta$ function. We see that even when a small dimension for the Hamiltonian is used, $N=5$, the simulation performs very well, provided that $\sigma$ is not too close to the boundary $\sigma=1$ of convergence of the Dirichlet series. The error of the approximation depends on the values of the parameters $a$ and $\sigma$. As $\sigma$ approaches unity, if $N$ is kept fixed, the error will become larger, as shown in the lower panel of Fig.~\ref{comparison}. We discuss more on that issue in Sec.V.
In the next section, we propose several implementations in particular physical systems.

\section{Physical Implementation}
The tridiagonal Hamiltonian introduced in the previous section arises in many different areas of physics. In particular, tight-binding Hamiltonians with engineered hopping rates and site energies have been extensively investigated in connection to the problem of perfect quantum transfer in quantum networks, and physical implementations based on e.g. spin chain models have been suggested (see, for instance, \cite{SC1,SC2,SC3,SC4} and references therein). Hence we do not aim  here to point out all the possible realizations of the proposed scheme. We will specifically focus on two possible physical implementations, namely in trapped ions and in optics.\\

\begin{figure}[tb]
\includegraphics[width=7cm]{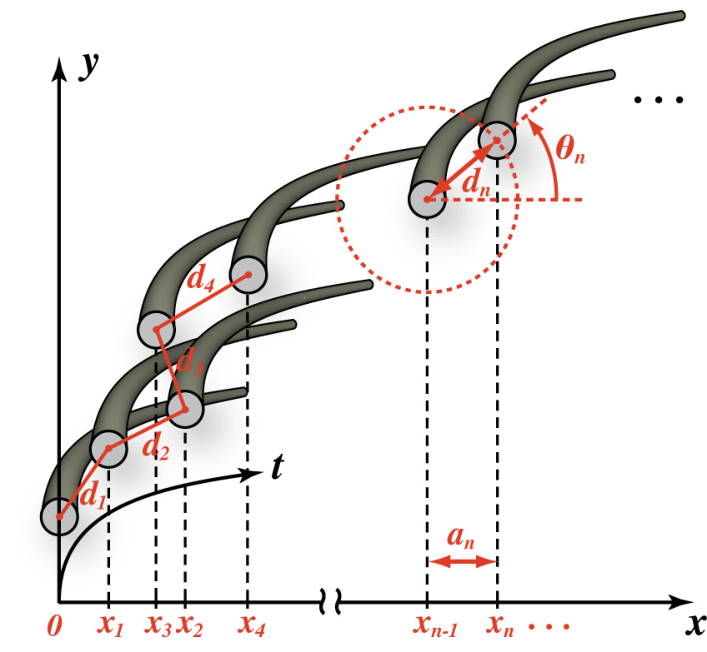}
\caption{Sketch of a possible physical implementation of the tridiagonal Hamiltonian $\mathbf{H}$ for the quantum simulation of RH  $\zeta$ function based on a chain of evanescently-coupled optical waveguides with circularly-curved optical axis.}
\label{implementation}
\end{figure}

A first possible  realization uses a linear chain of trapped ions. From an experimental point of view, trapped ions are one of the most advanced systems for quantum computation \cite{Trapped Ions Comp} and quantum simulation \cite{Trapped Ions Sim}. For the purpose of this work, we need to be able to experimentally construct a tridiagonal symmetric interaction matrix, with a control over each element of the Hamiltonian. This can be achieved, for instance, by implementing a nearest neighbour spin-spin interaction
\be\label{spin H}
\H = \tfrac12 \sum_{n=1}^{N-1} J_n (\sigma_n^{x}\sigma_{n+1}^{x}+\sigma_n^{y}\sigma_{n+1}^{y}) -\sum_{n=1}^{N} B_n \sigma_n^{z},
\ee
where $\sigma^{x,y,z}$ are the Pauli matrices, $J_n$ are the spin-spin coupling constants, and $B_n$ are individual couplings to an external (fictitious) magnetic field.
This Hamiltonian commutes with the excitation number operator, which means that the Hilbert space factorizes into subspaces with different number of excitations. If we consider the single-excitation subspace, it is easy to show that, in this basis, the Hamiltonian can be written as a tridiagonal matrix \cite{Kay}
\be\label{H-Kay}
\H = \left[\begin{array}{cccccc}
B_1 & J_1 & 0 & \ldots & 0 & 0 \\
J_1 & B_2 & J_2 &  \ldots & 0 & 0 \\
0 & J_2 & B_3 &  \ldots & 0 & 0 \\
\vdots & \vdots & \vdots &  \ddots & \vdots & \vdots \\
0 & 0 & 0 &  \ldots & B_{N-1} & J_{N-1} \\
0 & 0 & 0 &  \ldots & J_{N-1} & B_N
\end{array}\right],
\ee
where we neglect a common irrelevant term $-\sum_{n=1}^{N}B_n$ in all the diagonal elements. Traditionally, Hamiltonian \eqref{spin H} is implemented by using optical spin-dependent forces, induced by laser beams \cite{Trapped Ions Sim}. In this approach, off-resonant coupling to the motional sidebands is used in order to produce effective spin-spin interactions. Recently it has been shown that a general $J_{ij}$ coupling matrix can be implemented by using such a setup \cite{Monroe}.  As an example, for a spin chain made of  $N=5$ sites the normalized  values of couplings $J_n$ and local magnetic fields $B_n$, needed to realize the logarithmic spectrum \eqref{spectrum}, are given by Eq.~\eqref{explicit}. Note that the corresponding magnetic fields $B_n$ show  a slow monotonic increase, whereas the hopping rates $J_n$ change in the range of $ \pm 30 \%$ about a mean value. \par
A second possible physical implementation of the tridiagonal Hamiltonian is provided by light transport in a chain of evanescently-coupled optical waveguides with engineered coupling constants and propagation constants (see, for instance, \cite{cazzo} and references therein).  
In an array of  coupled optical waveguides, it is well-known that transport of discretized light is governed by a set of coupled-mode equations described by a tridiagonal Hamiltonian $\mathbf{H}$ of the same form as Eq.~\eqref{H-Kay}, where $N$ is the total number of waveguides, $B_n$ are the propagation constants of guided modes in the various waveguides of the array and $J_n$ are the coupling (hopping) constants between adjacent waveguides  ~\cite{cazzo,reviewuff}. In order to {\it independently} engineer the coupling $J_n$ and propagation constants $B_n$, it is convenient to circularly bend the axis of the waveguide array. A schematic of the optical waveguide setting that realizes the Hamiltonian $\mathbf{H}$  is shown in Fig.~\ref{implementation}. Axis bending basically introduces a mismatch between the propagation constants of adjacent waveguides due to different geometric paths. Such mismatch  can be properly tailored by controlling the radius of curvature $R$ of waveguides and the waveguide separation in the plane of axis bending (see, for instance, \cite{reviewuff}). On the other hand, the hopping rates $J_n$ between adjacent waveguides can be tailored by a proper choice of the waveguide separation. To appreciate the role of axis bending and to properly design the waveguide array, we recall that propagation of discretized light in the  bent waveguide array of Fig.~\ref{implementation} is governed by the coupled-mode equations~\cite{cazzo,reviewuff}:
\begin{equation}\label{waveguides}
i\frac{dc_n}{dt} = -J_{n-1} c_{n-1} - J_{n} c_{n+1} + \left(E_0 + \Delta E_n \right) c_n
\end{equation}
$(n=1,2,3,...,N$), where $c_n$ is the amplitude of light waves trapped in the $n$-th waveguide, $J_n$ is the hopping rate between waveguides $n$ and $n+1$ (with $J_0=J_N=0$), $E_0$ is the effective index of the fundamental mode of the individual waveguide with a straight optical axis, and $\Delta E_n = n_s x_n/(R \lambdabar)$ is the propagation constant detuning of waveguide $n$ induced by bending of the optical axis $t$, with $x_n$ the $x$-coordinate of waveguide $n$ [see Fig.~\ref{implementation}]. In the above Eq.~\eqref{waveguides}, $n_s$ is the refractive index of the substrate and $\lambdabar = \lambda/(2 \pi)$ with $\lambda$  the optical wavelength. For waveguide separation $d_n$, the hopping rate is given to an excellent accuracy by the exponential law $J_n = \kappa \exp(-\alpha d_{n})$, with $\kappa$ and $\alpha$ some constants depending on waveguide fabrication parameters that can be experimentally determined. The detuning between waveguide $n$ and $n-1$ is determined by the $x$-coordinate difference $a_n=x_n-x_{n-1}=d_n \cos \theta_n$ [see Fig.~\ref{implementation}], and thus can be geometrically controlled by selecting the proper value of $\theta_n$ in the interval $[0,\pi]$. To obtain the correlation function $\zeta(t)$ in the photonic structure of Fig.~\ref{implementation}, the left-boundary waveguide $n=1$ should be excited at the input plane $t=0$ by a light beam, e.g. using  butt-coupling by an optical fiber. The correlation RH function $\zeta(t)$ is then simply retrieved by monitoring the amplitude $c_1(t)$ of light that remains trapped in the $n=1$ waveguide of the lattice along the propagation distance $t$. 
It is worth noting that, with waveguide arrays manufactured by the femtosecond laser writing  technology \cite{writing}, a careful control of hopping rates and waveguide bending is nowadays possible. For example, tridiagonal Hamiltonian matrices with inhomogeneous hopping rates with $N$ up to a few tens have been recently demonstrated in femtosecond laser written wavegude chains \cite{palle}. Hence the waveguide design suggested in Fig.~\ref{implementation} is expected to be feasible with current waveguide fabrication technologies.

\section{Conclusions and discussion}

In this paper we have proposed an experimentally feasible method to simulate the Riemann-Hurwitz  $\zeta$ function by the autocorrelation function of a finite-dimensional quantum system initially prepared in a bare state. The proposal is based on an engineered one-dimensional tight-binding Hamiltonian, with specific conditions for the eigenvalues and eigenvectors, which can be implemented by e.g. trapped ion systems  or evanescently-coupled optical waveguide arrays. 
As compared to the recent proposal of Ref. \cite{Schleich} and based on the realization of an anharmonic quantum oscillator, our quantum analog simulator  uses a more feasible  tight-binding Hamiltonian and enables  a very simple preparation of  the system in a thermal (Riemannian) phase state. Like the proposal of Ref. \cite{Schleich}, our analog simulator uses the Dirichlet representation of the RH $\zeta$ function, and so it is regrettably not suited to simulate the more interesting behavior of the function along the critical line $1/2+i t$, where the nontrivial zeros of the Riemann function are located. A desirable extension of our results would be to synthesize a tight-binding Hamiltonian, which can simulate the behavior of the Riemann function along such a line. The starting point of such goal could be provided by the Riemann-Siegel formula \cite{Schleichbook}, which gives an approximation of the $\zeta$ function by a sum of two finite Dirichlet series which is valid along the critical line. It is envisaged that such two finite sums could be implemented by means of a finite-dimensional block-diagonal Hamiltonian, where each block simulates independently each of the two series in the Riemann-Siegel formula. In this proposal, however, the measurement of the autocorrelation function might be a more challenging task, and further investigation is required.\\
On a practical level, it should be mentioned that our proposal suffers from some limitations, which are similar to those discussed for other systems (see e.g. Ref. \cite{Schleich}). A first issue is related to the truncation of the Dirichlet series, which makes our approach unsuited to simulate the RH $\zeta$ function  near the line $\sigma=1$, where the convergence of the Dirichlet series becomes very slow and thus extremely large values of $N$ would be required.  In principle, the algorithm presented in Sec.II works well for small as well as large values of $N$ (e.g. $N$ up to 1000), however in any physical implementation $N$ is limited to a relatively small value because of technological issues. For example, in the photonic realization discussed in Sec.IV an upper limit to $N$ can be roughly estimated to be of a few tens (e.g. $N=50$). As $N$ is increased, tolerances in the values of $J_n$ and $B_n$ that ensure the logarithmic  spectrum \eqref{spectrum} of eigenvalues become more stringent and unrealistic. The limited number of $N$ available in any realistic implementation of $\mathbf{H}$ poses a restriction on the  domain of the RH $\zeta$ function that we can access in an experiment owing to the slow convergence of the series \eqref{Dirichlet} as $\sigma \rightarrow 1^+$. For a given value of $\sigma$, the error introduced by truncation of the series \eqref{Dirichlet} remains at an acceptable small level provided that $N$ is larger than a minimum value $N_{\min}$, that rapidly increases as  $\sigma \rightarrow 1^+$. To estimate $N_{\min}$, let us consider as an example the Riemann function, i.e. the case $a=1$, and let us consider the asymptotic behavior of the error
\be\label{err1}
\epsilon = \zeta(s,1) - \sum_{n=0}^{N-1}\frac{1}{(n+1)^s}= \sum_{n=N}^{\infty}\frac{1}{(n+1)^s},
\ee
as ${\rm Re}(s)=\sigma \rightarrow 1^+$. The asymptotic form of $\epsilon$ for the Riemann function is given by Theorem 4.11 in Titchmarsh \cite{Tich}. If we  consider for the sake of clearness the $t=0$ case, one has  \cite{Tich}
\begin{equation}\label{err2}
\epsilon(N,\sigma)=-\frac{N^{1-\sigma}}{1-\sigma}+ O \left( \frac{1}{N}^{\sigma} \right).
\end{equation}
Note that, as $\sigma \rightarrow 1^+$, the first term on the right hand side in Eq.~\eqref{err2} gets singular, and to keep the error small one should take $N$ large enough to counteract the singular term. For a given value of $\sigma$ close to (but larger than) one, an estimate of the truncation index $N$ that ensures a small error $| \epsilon| \ll 1$ is readily obtained from Eq.~\eqref{err2} and reads $N \gg N_{\min}(\sigma)$, where
\begin{equation}
N_{\min}(\sigma)= \left( \sigma-1 \right)^{-\frac{1}{\sigma-1}}
\end{equation}
The behavior of $N_{\min}$ versus $\sigma$ shows a steep increase below $\sigma \sim 1.3$. For example, one has $N_{\min} \simeq 4, 55,3125$ for $\sigma=1.5, 1.3, 1.2$, respectively.  This shows that an accurate estimate of the Riemann function can be realized in practice for $\sigma$ larger than $ \sigma_{\min} \simeq 1.3$, and that for $\sigma>1.5$ the estimation is very accurate even for few lattice sites $N$.
Another limitation of our proposed scheme  in the estimation of the RH correlation function is decoherence. Obviously, our analysis assumes that the quantum evolution of the system is coherent, i.e. we neglected interaction with the environment. This requires that the observation time $t$ is smaller than the coherence time $t_{\coh}$. This limitation will depend on the specific physical implementation of the quantum simulator. In trapped ions coherence time varies in the range of milliseconds to minutes, depending on the qubit realization and the ions in use. On the other hand, in the case of the optical-waveguides realization decoherence is irrelevant and can be ignored, though the maximum value of $t$ is still limited by the sample length. The two limitations for the parameters $t$ and $\sigma$ discussed above, arising from decoherence and from series truncation, are illustrated in Fig.~\ref{Region}.
\acknowledgments

This work was supported by the Fondazione Cariplo (Grant No. 2011-0338).


\end{document}